# Dispersion relation of nutation surface spin waves in ferromagnets

Mikhail Cherkasskii[1, *], Michael Farle[2,3], and Anna Semisalova[2]

[1] *Department of General Physics 1, St. Petersburg State University, St. Petersburg, 199034, Russia*
[2] *Faculty of Physics and Center for Nanointegration (CENIDE), University of Duisburg-Essen, Duisburg, 47057, Germany*
[3] *Kirensky Institute of Physics, Federal Research Center KSC SB RAS, Krasnoyarsk, 660036, Russia*

\* Corresponding author: macherkasskii@hotmail.com

Inertia effects in magnetization dynamics are theoretically shown to result in a different type of spin waves, i.e. nutation surface spin waves, which propagate at terahertz frequencies in in-plane magnetized ferromagnetic thin films. Considering the magnetostatic limit, i.e. neglecting exchange coupling, we calculate dispersion relation and group velocity, which we find to be slower than the velocity of conventional (precession) spin waves. In addition, we find that the nutation surface spin waves are backward spin waves. Furthermore, we show that inertia causes a decrease of the frequency of the precession spin waves, namely magnetostatic surface spin waves and backward volume magnetostatic spin waves. The magnitude of the decrease depends on the magnetic properties of the film and its geometry.

## I. INTRODUCTION

From the classical point of view, spin waves are collective excitations of magnetically ordered materials, that is waves of precession of the magnetization [1,2], for example in thin magnetic films [3-5], layered magnetic structures [6,7], periodic magnetic crystals [8,9], and nanometer-sized structures [10]. These waves exhibit typical linear [1,2] and nonlinear wave effects [11-13], such as excitation [14,15], propagation [16-18], reflection [19,20], and interference [21,22] in the first case, and self-focusing [23-26], formation of envelope solitons [27,28], chaotic behavior [29], as well as parametric three- and four-waves processes [30,31] in the non-linear case.

Recently, it has been theoretically and experimentally demonstrated that the effects of inertia of magnetization should be considered in the full description of spin dynamics at pico- and femtosecond timescales [32-40]. The nutation motion of magnetization is a manifestation of inertia of the magnetic moments. A rigorous derivation including inertia in the Landau-Lifshitz-Gilbert equation was carried out by Mondal et al. in the Dirac-Kohn-Sham framework [33,34]. A relation between the Gilbert damping and the inertial characteristic time was investigated in Ref. [32]. In another approach M.-C. Ciornei et al. confirmed that inertia is responsible for nutation, and that this motion is superimposed on the precession of magnetization [38]. The influence of nutation on the dynamic susceptibility was analytically [41] and numerically [42] studied.

Despite these theoretical advances, the experimental study of inertial spin dynamics has only begun. Following the indirect observation of inertial magnetization dynamics in $Ni_{79}Fe_{21}$ and Co films [43], direct experimental confirmation of nutation resonance was reported by Neeraj et al. [37].

In this paper, we predict an additional effect, that is the emergence of propagating nutation surface spin waves (NSSW) in the dipole–dipole coupling limit, and the transformation of conventional precession waves to precession-nutation spin waves. We derive dispersion relation of NSSW and calculate the spectral shift of precession-nutation spin waves with respect to precession spin waves. The emergence of nutation waves due to exchange coupling rather then dipolar interaction has been proposed by Makhfudz et al. [44] recently.

In general, the following interactions must be taken into account to describe the dynamics of spin waves: Zeeman, spin-orbit, exchange, and dipole-dipole interactions. The phase shift between precessing magnetic moments propagates as a spin wave through the ferromagnet because of dipole-dipole or exchange coupling (Fig. 1(a)). Magnetic inertia effects, which are expected to contribute to dynamics of spin waves, originate from spin-orbit coupling (coupling of the spins to the lattice via the orbital moment). In magnetization dynamics, this relativistic effect is considered with different orders of approximation. In the lowest order, one obtains the Gilbert damping of magnetization precession and the gyromagnetic ratio, i.e., the relation between angular momentum and spin. In higher order approximations, magnetic inertia appears [33,34,45], and the gyromagnetic ratio must be generalized, which leads to nutation motion of magnetic moments superimposed on their precession. Taking inertia into account one finds that the deviation of localized moments will propagate through the spin system in the form of *both* precession and nutation motions, i.e. in ferromagnetic materials one needs to add to all "conventional" spin wave modes a high frequency wave-like motion with small amplitude caused by inertia. Additionally, waves having predominantly inertial nature appear in ferromagnetic thin films, which we call here *nutation surface spin waves*. Since these waves have terahertz frequencies (compared to typically GHz frequencies of other spin wave modes), they can be plotted as a small deviation on top of a "frozen" precession motion (Fig. 1(b)).

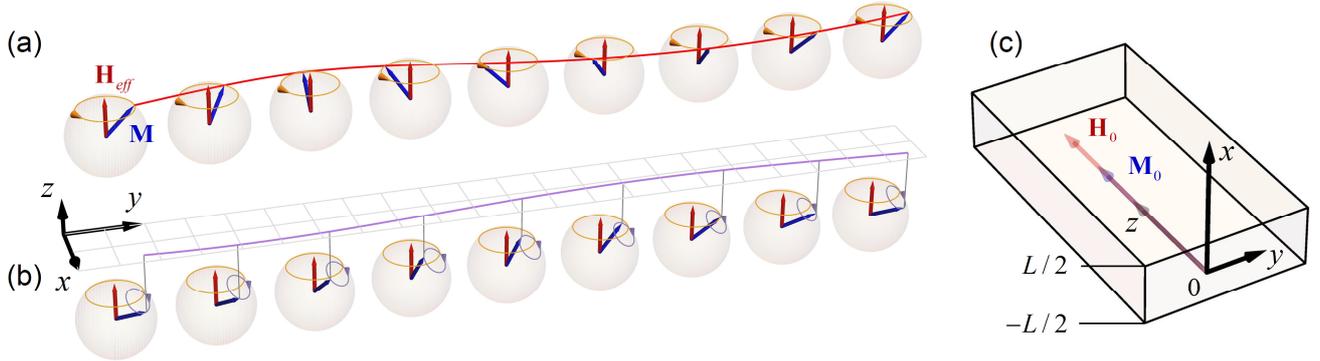

FIG. 1. (Color online) (a) The precession spin wave without inertia (red curve). The blue arrow indicates the motion of the magnetization **M** in a film. (b) The nutation surface spin wave (purple curve) with a frequency considerably higher than in (a) plotted with small blue circles on top of the "frozen" precession motion. (c) Coordinate system of the ferromagnetic film with thickness $L$, magnetization $\mathbf{M}_0$ and applied magnetic field $\mathbf{H}_0$.

In our calculation, we work in the dipole-dipole coupling limit, which allows us to use a magnetostatic approach in which Maxwell's equations are transformed into the Walker equation [4]. To obtain the dispersion relation including inertia, we use the dynamic susceptibility derived from the inertial Landau-Lifshitz-Gilbert (ILLG) equation [41] and substitute the result into the Walker equation.

In this paper, we consider waves propagating in thin ferromagnetic films magnetized in-plane by an external magnetic field. We focus on two particular configurations: (A) waves propagating perpendicular to the external magnetic field $\mathbf{H}_0$ (see Fig. 1(c), y-axis), and (B) waves propagating along $\mathbf{H}_0$ (Fig. 1(c), z-axis). The latter case (B) corresponds to backward volume magnetostatic spin waves (BVMSW), when only precession is taken into account, and denoted as n-BVMSW when precession-nutation case is considered. Similarly, for perpendicular configuration (A) we distinguish in our nomenclature between magnetostatic surface spin waves (MSSW), i.e. in other words Damon-Eshbach mode, when inertia is neglected, and these waves as n-MSSW when inertia is included. Finally, for perpendicular configuration our calculation predicts a different type of waves – nutation surface spin waves. Note that the n-MSSW are conventional precession spin waves whose dispersion is modified by inertia effects due to nutation (Fig. 2(b) (bottom) and (e)), while nutation surface spin waves are solely due to nutation which occurs on top of the background precession.

## II. DISPERSION EQUATIONS AND WAVE CHARACTERISTICS

The ferromagnetic film, magnetic field and coordinate system are shown in Fig. 1(c). The film with thickness $L$ is placed in an external magnetic field $\mathbf{H}_0$ strong enough to saturate the magnetization of the film. We assume that the exciting magnetic field is small $|\mathbf{h}| \ll |\mathbf{H}_0|$, and the static magnetization vector $\mathbf{M}_0$ and external magnetic field $\mathbf{H}_0$ are aligned.

Maxwell's equations in magnetostatics are written as

$$\nabla \times \mathbf{h} = 0, \quad (1)$$
$$\nabla \cdot (\mathbf{h} + \mathbf{m}) = 0, \quad (2)$$

where $\mathbf{m}$ is the response of the magnetization to the small driving magnetic field. Equation (1) allows to introduce the magnetic potential using $\mathbf{h} = \nabla \psi$, substitute this potential into equation (2) and obtain Walker's equation

$$(1+\chi)\left(\frac{\partial^2 \psi}{\partial x^2} + \frac{\partial^2 \psi}{\partial y^2}\right) + \frac{\partial^2 \psi}{\partial z^2} = 0, \quad (3)$$

where $\chi$ is the diagonal component of the dynamic susceptibility tensor (see the Supplemental Material [46]). The potential obeys Laplace's equation outside of the film

$$\nabla^2 \psi = 0. \quad (4)$$

Following the ansatz and interface conditions from Ref. [4], the characteristic equations from Walker's and Laplace's equations are obtained

$$(1+\chi)\left(\left(k_x^i\right)^2 + k_y^2\right) + k_z^2 = 0,$$
$$\left(k_x^e\right)^2 - k_y^2 - k_z^2 = 0, \quad (5)$$

where $k_x^{i,e}$, $k_y$ and $k_z$ denote wavenumbers along the $x$, $y$, $z$ axes while superscripts indicate internal $(i)$ and external $(e)$ wavenumbers with respect to the film boundaries. The characteristic equations (5) determine the allowed limit of wavenumbers of propagating spin waves. To investigate propagating waves one employs the real parts of the dynamic susceptibility in equation (6)

$$(k_x^e)^2 + 2k_x^e k_x^i (1+\chi')\cot(k_x^i L) \\ -(k_x^i)^2 (1+\chi')^2 - k_y^2 \chi_a'^2 = 0, \quad (6)$$

where $\chi'$ is the real dispersive part of $\chi$, and $\chi_a'$ is the real dispersive part of the anti-diagonal component of the dynamic susceptibility tensor. The effect of inertia of the magnetization is introduced by the dynamic susceptibility deduced from the ILLG equation. The detailed derivation for a Cartesian coordinate system can be found in the Supplemental Material [46]. In the following sections, we focus on the dispersion relations for spin waves propagating in perpendicular (A) and parallel (B) direction to the magnetic field.

## A. Perpendicular configuration

For spin waves propagating in the perpendicular direction to the external magnetic field, equation (6) becomes

$$(1+\chi')(1+\coth(k_y L)) - \frac{\chi_a'^2 - \chi'^2}{2} = 0, \quad (7)$$

since $k_z$ should be equal to zero in this configuration. The substitution of the susceptibility expressions (equations (S7)-(S11) from Ref. [46]) into (7) allows us to calculate the dispersion relation between frequency and wavenumber $k_y$. This substitution leads to the bi-quartic equation

$$A_s \omega^8 + B_s \omega^6 + C_s \omega^4 + D_s \omega^2 + E_s = 0, \quad (8)$$

where

$$A_s = 2\alpha^4 \tau^4 (1+\coth(k_y L)), \quad (9)$$

$$B_s = 2\alpha^2 \tau^2 (-2 + 2\alpha^2 - \alpha\tau(4\omega_H + \omega_M)) \\ \times (1+\coth(k_y L)), \quad (10)$$

$$C_s = 2 + 2\alpha^4 + 2\alpha\tau(4\omega_H + \omega_M) \\ - 2\alpha^3 \tau(4\omega_H + \omega_M) \\ + 4\alpha^2 + \alpha^2 \tau^2 (12\omega_H^2 + 6\omega_H \omega_M + \omega_M^2) \\ + 2\big[\{1 + \alpha\tau(4\omega_H + \omega_M) \\ -\alpha^3\tau(4\omega_H + \omega_M) \\ + 2\alpha^2 + 3\alpha^2\tau^2\omega_H(2\omega_H + \omega_M) + \alpha^4\big]\coth(k_y L), \quad (11)$$

$$D_s = -8\alpha\tau\omega_H^3 - \omega_M^2 + \omega_H^2(-4 + 4\alpha^2 - 6\alpha\tau\omega_M) \\ - 2\omega_H\omega_M(1 - \alpha^2 + \alpha\tau\omega_M) \\ - 2\omega_H\big[4\alpha\tau\omega_H^2 + \omega_M - \alpha^2\omega_M \\ + \omega_H(2 - 2\alpha^2 + 3\alpha\tau\omega_M)\big]\coth(k_y L), \quad (12)$$

$$E_s = \omega_H^2 \big[2\omega_H^2 + 2\omega_H\omega_M + \omega_M^2 \\ + 2\omega_H(\omega_H + \omega_M)\coth(k_y L)\big]. \quad (13)$$

$$\omega_H = |\gamma|\mu_0 H_0, \\ \omega_M = |\gamma|\mu_0 M_0. \quad (14)$$

Here, $\gamma$ is the gyromagnetic ratio, $M_0$ is the magnetization at saturation, $\alpha$ is the dimensionless Gilbert damping parameter, $\tau$ is the inertial relaxation time. Note that $\tau$ is inversely proportional to $\alpha$ [45], therefore characteristic time $\alpha\tau$ was proposed to describe inertial behavior [40,42]. We employ Ferrari's method for finding the solutions of equation (8), and introduce the notation:

$$a_s = \frac{C_s}{A_s} - \frac{3B_s^2}{8A_s^2}, \\ b_s = -\frac{B_s C_s}{2A_s^2} + \frac{B_s^3}{8A_s^3} + \frac{D_s}{A_s}, \quad (15) \\ c_s = \frac{B_s^2 C_s}{16A_s^3} - \frac{3B_s^4}{256A_s^4} - \frac{B_s D_s}{4A_s^2} + \frac{E_s}{A_s}.$$

In Ferrari's method, one determines the root of the nested depressed cubic equation written here as:

$$y_s = -\frac{5a_s}{6} + U_s + V_s, \quad (16)$$

where

$$U_s = \sqrt[3]{-\sqrt{\frac{P_s^3}{27} + \frac{Q_s^2}{4}} - \frac{Q_s}{2}}, \\ V_s = -\frac{P_s}{3U_s}, \quad (17) \\ P_s = -\frac{a_s^2}{12} - c_s, \\ Q_s = \frac{1}{3}a_s c_s - \frac{a_s^3}{108} - \frac{b_s^2}{8}.$$

The bi-quartic equation (8) has four roots describing the relationship of frequency and wavenumber, i.e. different dispersion branches. One branch corresponds to zero wavenumber, i.e. uniform precession and ferromagnetic resonance (FMR). The second dispersion branch resembles MSSW (Fig. 2(b) (bottom) and (e)) and merges with it at $\alpha\tau = 0$ when inertia is neglected. We suggest to denote this branch n-MSSW.

The n-MSSW branch is given by the expression

$$\omega_{MSSW}^n = \left(-\frac{B_s}{4A_s} - \frac{\sqrt{a_s + 2y_s}}{2} \\ + \frac{1}{2}\sqrt{-3a_s - 2y_s + \frac{2b_s}{\sqrt{a_s + 2y_s}}}\right)^{1/2}. \quad (18)$$

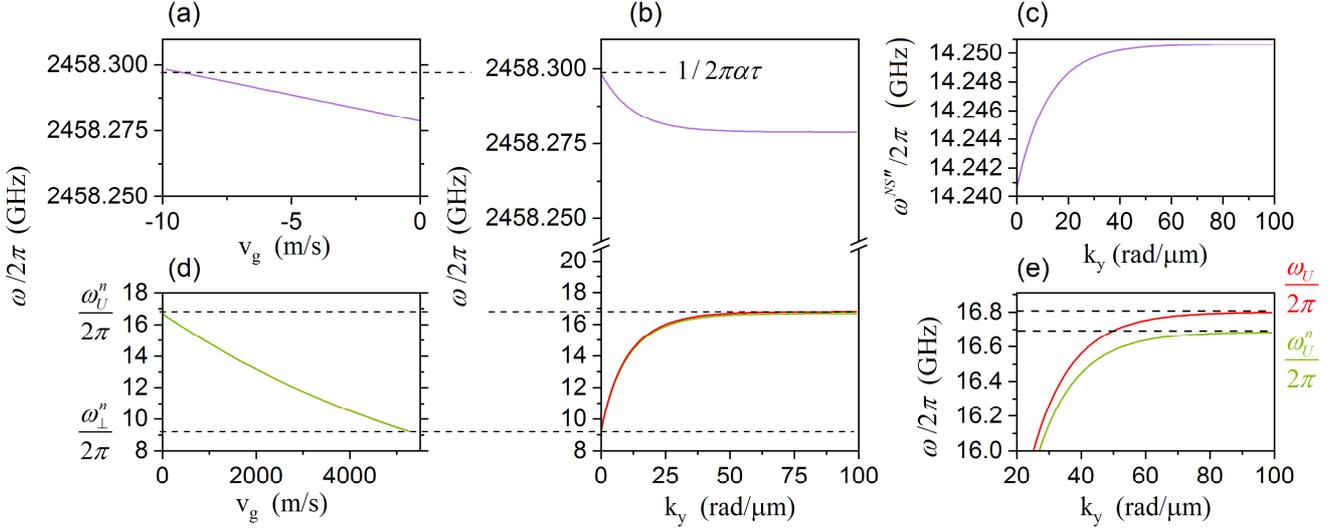

FIG. 2. (Color online) (a) The group velocity of nutation surface spin waves. (b) The dispersion branches of nutation surface spin waves (purple curves) in terahertz range, MSSW (red curves) and n-MSSW (green curves) in microwave range. (c) The inherent losses of nutation surface spin waves. (d) The group velocity of the n-MSSW. (e) The dispersion branches of MSSW (red curves) and n-MSSW (green curves) in magnified scale. The parameters for the calculation of precession-nutation waves (n-MSSW) are $\mu_0 M_0 = 1$ T, $\mu_0 H_0 = 100$ mT, $\alpha = 0.0065$, and $\tau = 10^{-11}$ s and for MSSW are the same parameters except $\alpha = 0$ and $\tau = 0$.

The frequency in (18) is real, hence n-MSSW propagate as sinusoidal waves. There is a spectral red-shift between the n-MSSW and MSSW branches, and in the following we study in detail the shift of the spectrum limits. The upper and lower limits of the spectrum are shifted down differently. Without nutation the dispersion branch of MSSW exists in the frequency range $\omega_\perp < \omega < \omega_U$, where $\omega_U = \omega_H + \omega_M / 2$, and $\omega_\perp = \sqrt{\omega_H (\omega_H + \omega_M)}$. The upper spectrum limit of the n-MSSW $\omega_U^n$ can be calculated with the expression (18) at $k_y \to \infty$ that yields $\coth(k_y L) = 1$. For instance, we calculate the difference between upper limits for the following parameters $\mu_0 M_0 = 1$ T, $\mu_0 H_0 = 100$ mT, $\alpha = 0.0065$, and $\tau = 10^{-11}$ s. The difference is 115 MHz or 0.7%, and it can be clearly seen in Fig 2(e). The decrease of frequencies of n-MSSW is caused by nutation, which is a rotation of magnetization in the opposite direction compared to the precession [46].

Due to the fact that the lower spectrum limit of the MSSW corresponds to upper spectrum limit of the BVMSW, we discuss the spectral red-shift of the lower limit of n-MSSW in section II (B).

The third and fourth dispersion branches following from (8) are complex conjugates, their frequency at zero wavenumber corresponds to nutation resonance frequency. In case of $\alpha\tau = 0$, these branches vanish. We assign these two branches to a different type of spin waves, NSSWs, which represent a wave of propagating nutation, and we plot the real and imaginary parts of dispersion curves in Fig. 2(b) and (c).

The real part of the branch describing propagation of the NSSW is determined by

$$\omega^{NS\prime} = \frac{w_1 + w_2}{2}. \quad (19)$$

The imaginary part corresponding to the inherent losses of NSSW is written as

$$\omega^{NS\prime\prime} = \frac{w_1 - w_2}{2i}, \quad (20)$$

where

$$w_1 = \left( -\frac{B_s}{4A_s} + \frac{\sqrt{a_s + 2y_s}}{2} \right.$$
$$\left. -\frac{1}{2}\sqrt{-3a_s - 2y_s - \frac{2b_s}{\sqrt{a_s + 2y_s}}} \right)^{1/2}, \quad (21)$$

$$w_2 = \left( -\frac{B_s}{4A_s} + \frac{\sqrt{a_s + 2y_s}}{2} \right.$$
$$\left. +\frac{1}{2}\sqrt{-3a_s - 2y_s - \frac{2b_s}{\sqrt{a_s + 2y_s}}} \right)^{1/2}. \quad (22)$$

The NSSW have inherent losses, since these waves exist only if $\alpha\tau \neq 0$. This fact is the direct consequence of their inertial nature – these waves do not exist, if one neglects inertia and damping.

The magnetostatic potential of spin waves is concentrated close to the surface of the ferromagnetic film if the magnetic field is applied perpendicular to the wavevector, hence in this configuration conventional spin waves and nutation spin waves are surface waves. However, NSSW have lower group velocity than the precession waves that can be seen from Figs. 2(a) and (d). The group velocity is calculated using the standard expression $v_g = \partial \omega / \partial k_y$. The negative value of the group velocity means that NSSW are backward waves.

The highest frequency of the NSSW is the nutation resonance frequency $1/\alpha\tau$, derived earlier in Ref. [41].

## B. Parallel configuration

If one considers waves propagating parallel to the external magnetic field direction, $k_y = 0$, and equation (6) can be simplified as

$$1 + \frac{\chi'}{2} + \sqrt{1+\chi'} \coth\left(\frac{k_z L}{\sqrt{1+\chi'}}\right) = 0. \tag{23}$$

We substitute the susceptibility expressions (S7)-(S11) from Ref. [46] and employ numerical damped Newton's method for finding the roots of the algebraic equation (23) to calculate the relations between frequency and wavenumber in both precession and precession-nutation cases. These relations demonstrate a set of dispersion branches, and the first three branches are plotted in Fig. 3.

It is clearly seen from Fig. 3 that the dispersion of n-BVMSW is shifted relatively to the BVMSW. To investigate this, we compare the spectral limits of the precession and precession-nutation waves. Note that the volume waves exist only in the frequency range where $1+\chi' \le 0$. This condition is a consequence of equation (5), since propagating waves have real wavenumber $k_x^i \in R$. Therefore, the spectrum limits are defined by

$$1 + \chi' = 0, \tag{24}$$

and in the non-damping and non-inertia case this equation determines the frequency range $\omega_H < \omega < \omega_\perp$, where $\omega_\perp = \sqrt{\omega_H (\omega_H + \omega_M)}$. If one takes nutation into account, the numerator of equation (24) allows us to find the frequency range. This numerator can be written as

$$A_v \omega^8 + B_v \omega^6 + C_v \omega^4 + D_v \omega^2 + E_v = 0, \tag{25}$$

where

$$\begin{aligned}
A_v &= \alpha^4 \tau^4, \\
B_v &= \alpha^2 \tau^2 \left(-2 + 2\alpha^2 - \alpha\tau(4\omega_H + \omega_M)\right), \\
C_v &= 1 + \alpha\tau(4\omega_H + \omega_M) \\
&\quad + 2\alpha^2 + 3\alpha^2\tau^2 \omega_H (2\omega_H + \omega_M) \\
&\quad - \alpha^3 \tau (4\omega_H + \omega_M) \\
&\quad + \alpha^4, \\
D_v &= -\omega_H \big[ 4\alpha\tau\omega_H^2 + \omega_M - \alpha^2 \omega_M \\
&\quad + \omega_H (2 - 2\alpha^2 + 3\alpha\tau\omega_M) \big], \\
E_v &= \omega_H^3 (\omega_H + \omega_M).
\end{aligned} \tag{26}$$

We repeat the procedure for finding the solutions using Ferrari's method. The spectrum limits of n-BVMSW must be found in the same way as provided in (15)-(17) with the corresponding replacement of variables, i.e. the subscript $s$ is replaced by $v$, which denotes volume waves. Thus, the upper limit of the spectrum is determined by the expression

$$\omega_\perp^n = \left( -\frac{B_v}{4A_v} - \frac{\sqrt{a_v + 2y_v}}{2} \right.$$
$$\left. + \frac{1}{2}\sqrt{-3a_v - 2y_v + \frac{2b_v}{\sqrt{a_v + 2y_v}}} \right)^{1/2}. \tag{27}$$

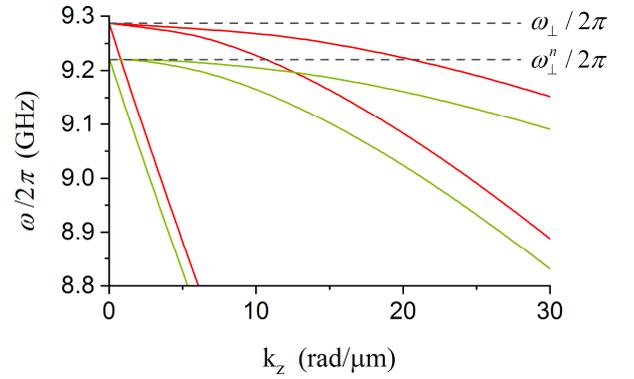

FIG. 3. (Color online) Dispersion branches for n-BVMSW (green curves) and for BVMSW (red curves). The parameters of the calculation for precession-nutation waves (n-BVMSW) are $\mu_0 M_0 = 1$ T, $\mu_0 H_0 = 100$ mT, $\alpha = 0.0065$, and $\tau = 10^{-11}$ s. The parameters for BVMSW are the same except $\alpha = 0$ and $\tau = 0$.

This upper spectrum limit of the n-BVMSW equals the lower spectrum limit of n-MSSW. The lower limit of the n-BVMSW is written as

$$\omega_H^n = \left( -\frac{B_v}{4A_v} - \frac{\sqrt{a_v + 2y_v}}{2} \right.$$
$$\left. - \frac{1}{2}\sqrt{-3a_v - 2y_v + \frac{2b_v}{\sqrt{a_v + 2y_v}}} \right)^{1/2}. \tag{28}$$

Expression (28) corresponds to the FMR frequency taking damping and nutation into account. For material parameters given in the previous section (Fig. 2) and in Fig. 3, the lower limit is shifted down by $3.1\,\text{MHz}$ compared to the case without damping and nutation effects $\omega_H = |\gamma|\mu_0 H_0$, and the upper limit decreases by $64\,\text{MHz}$. Thus, similarly to the perpendicular configuration of the wave vector and magnetization, the main effect of magnetization nutation for BVMSW is the red-shift of their dispersion branches.

## III. CONCLUSION

We theoretically predict the emergence of nutation surface spin waves due to magnetization inertia in the dipolar coupling limit for in-plane magnetized ferromagnetic thin films, propagating perpendicular to the direction of the external magnetic field. These waves are backward waves and propagate at terahertz frequencies with a group velocity lower than the velocity of conventional spin waves. Inertia leads to a red-shift of precession-nutation spin waves compared to precession spin waves. The upper spectral limit of the dispersion branches of the precession-nutation waves undergoes a greater shift than the lower spectral limit.

## ACKNOWLEDGMENTS

We acknowledge partial funding by Deutsche Forschungsgemeinschaft (DFG, German Research Foundation) – Project No. 392402498 (SE 2853/1-1) and Project No. 405553726 CRC/TRR 270, and by the government of the Russian Federation (Agreement No. 075-15-2019-1886). We thank J.-E. Wegrowe, U. Nowak and R. Mondal for valuable comments and helpful discussions.

# Supplemental Material to "Dispersion relation of nutation surface spin waves in ferromagnets"

Mikhail Cherkasskii[1, *], Michael Farle[2,3], and Anna Semisalova[2]

[1] Department of General Physics 1, St. Petersburg State University, St. Petersburg, 199034, Russia
[2] Faculty of Physics and Center for Nanointegration (CENIDE), University of Duisburg-Essen, Duisburg, 47057, Germany
[3] Kirensky Institute of Physics, Federal Research Center KSC SB RAS, Krasnoyarsk, 660036, Russia

* Corresponding author: macherkasskii@hotmail.com

The dynamic susceptibility in Cartesian coordinates is derived based on the inertial Landau-Lifshitz-Gilbert (ILLG) equation given by [1]

$$\frac{d\mathbf{M}}{dt} = -|\gamma|\mathbf{M} \times \left[ \mathbf{H}_{eff} - \frac{\alpha}{|\gamma|M_0}\left(\frac{d\mathbf{M}}{dt} + \tau \frac{d^2\mathbf{M}}{dt^2}\right)\right], \quad (S1)$$

where $\gamma$ is the gyromagnetic ratio, $\mathbf{M}$ is the magnetization vector, $M_0$ is the magnetization at saturation, $\alpha$ is the Gilbert damping, $\tau$ is the inertial relaxation time, and $\mathbf{H}_{eff}$ is the effective magnetic field. Since $\tau$ is inversely proportional to the damping parameter $\alpha$ [2], a characteristic time $\alpha\tau$ was proposed to describe inertia [3,4]. The spin-orbit interaction is the reason for Gilbert damping and magnetic inertia, which are described by the terms $\mathbf{M} \times d\mathbf{M}/dt$, and $\mathbf{M} \times d^2\mathbf{M}/dt^2$ and corresponding coefficients. Note that the susceptibility was derived for circular variables of magnetization and magnetic field in Ref. [5].

We assume that the ferromagnet is magnetically saturated by a uniform magnetic field $\mathbf{H}_0$ acting along the z-axis. The time-varying driving field $\mathbf{h}$ is superimposed on $\mathbf{H}_0$ and the linearization of ILLG can be performed since $|\mathbf{h}| \ll |\mathbf{H}_0|$. With the magnetization parallel to $\mathbf{H}_0$ we write the magnetization and magnetic field in the generalized form using the Fourier transformation

$$\mathbf{M}(t) = M_0 \hat{z} + \frac{1}{\sqrt{2\pi}} \int_{-\infty}^{\infty} d\omega' \, \mathbf{m}(\omega') e^{i\omega't}, \quad (S2)$$

$$\mathbf{H}_{eff}(t) = H_0 \hat{z} + \frac{1}{\sqrt{2\pi}} \int_{-\infty}^{\infty} d\omega' \, \mathbf{h}(\omega') e^{i\omega't}, \quad (S3)$$

where $\hat{z}$ is the unit vector along the z-axis. After several transformations the ILLG equation can be simplified to

$$i\omega \mathbf{m}(\omega) = -|\gamma|M_0 \hat{z} \times \mathbf{h}(\omega) + \\ + |\gamma|H_0 \hat{z} \times \mathbf{m}(\omega) + i\alpha\omega \hat{z} \times \mathbf{m}(\omega) \\ - \alpha\tau\omega^2 \hat{z} \times \mathbf{m}(\omega). \quad (S4)$$

By projecting to Cartesian coordinates one obtains

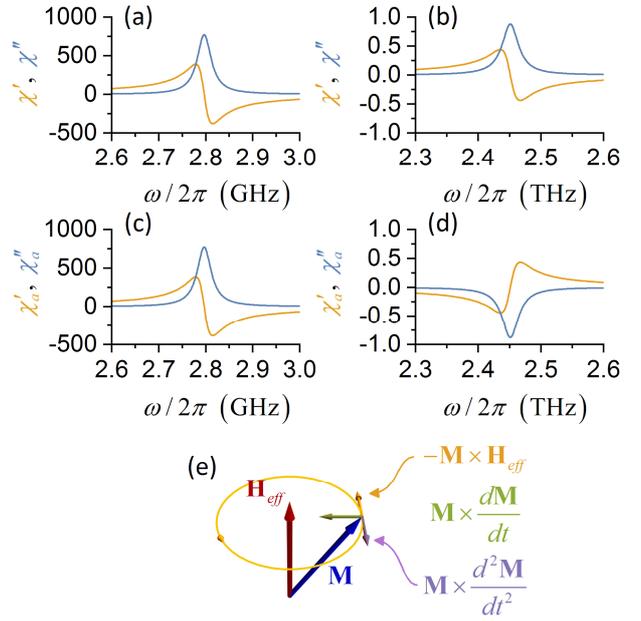

FIG. S1. (Color online) (a) The dispersive part of the susceptibility $\chi'$ (orange curve), and the dissipative part of the susceptibility $\chi''$ (blue curve). (b) The susceptibility components $\chi'$ and $\chi''$ presented in terahertz range. The panels (c) and (d) showing the antidiagonal susceptibility $\chi_a$. The calculation was performed for $|\gamma|/(2\pi) = 28$ GHz T$^{-1}$, $\mu_0 M_0 = 1$ T, $\mu_0 H_0 = 100$ mT, $\alpha = 0.0065$, and $\tau = 10^{-11}$ s. (e) The orientation of vectors in the ILLG equation.

$$\mathbf{m} = \hat{\chi}\mathbf{h},$$

$$\hat{\chi} = \begin{bmatrix} \chi & i\chi_a & 0 \\ -i\chi_a & \chi & 0 \\ 0 & 0 & 0 \end{bmatrix}, \quad (S5)$$

where

$$\chi = \frac{-\alpha\tau\omega_M\omega^2 + i\alpha\omega_M\omega + \omega_H\omega_M}{D},$$

$$\chi_a = \frac{\omega_M\omega}{D},$$

$$\begin{aligned}D = &\ \alpha^2\tau^2\omega^4 - 2i\alpha^2\tau\omega^3 \\ &- (1+\alpha^2+2\alpha\tau\omega_H)\omega^2 \\ &+ 2i\alpha\omega_H\omega + \omega_H^2\end{aligned} \quad \text{(S6)}$$

We introduced the following notations (in SI units): $\omega_H = |\gamma|\mu_0 H_0$, and $\omega_M = |\gamma|\mu_0 M_0$. The dynamic susceptibility tensor $\hat{\chi}$ is related to susceptibility in circular variables as $\chi_\pm = \chi \pm \chi_a$. Note that the precession resonance occurs for $\chi_+$ corresponding right-hand rotation, i.e. positive polarization, while nutation resonance has negative polarization. This can be also verified considering the vector orientations of ILLG equation (Fig. S1(e)). In the low damping limit, the vector orientation of $d\mathbf{M}/dt$ is identical to $\mathbf{M}\times\mathbf{H}_0$, and it is easy to find the orientation of $d^2\mathbf{M}/dt^2$ using this rule. It is clearly seen that the directions of precession torque $-|\gamma|\mathbf{M}\times\mathbf{H}_{\text{eff}}$, and nutation $(\alpha\tau/M_0)\mathbf{M}\times d^2\mathbf{M}/dt^2$ are opposite, and this corresponds to reversed polarizations of ferromagnetic and nutation resonances.

The dispersive and dissipative parts of the susceptibility can be separated with $\chi = \chi' - i\chi''$, $\chi_a = \chi_a' - i\chi_a''$, then

$$\begin{aligned}\chi' = \frac{1}{N}\big[&-\alpha^3\tau^3\omega_M\omega^6 \\ &+ (\alpha\tau\omega_M - \alpha^3\tau\omega_M + 3\alpha^2\tau^2\omega_H\omega_M)\omega^4 \\ &+ (-\omega_H\omega_M + \alpha^2\omega_H\omega_M - 3\alpha\tau\omega_H^2\omega_M)\omega^2 \\ &+ \omega_H^3\omega_M\big],\end{aligned} \quad \text{(S7)}$$

$$\begin{aligned}\chi'' = \frac{\omega}{N}\big[&\alpha^3\tau^2\omega_M\omega^4 \\ &+ (\alpha\omega_M + \alpha^3\omega_M - 2\alpha^2\tau\omega_H\omega_M)\omega^2 \\ &+ \alpha\omega_H^2\omega_M\big],\end{aligned} \quad \text{(S8)}$$

$$\begin{aligned}\chi_a' = \frac{\omega}{N}\big[&\alpha^2\tau^2\omega_M\omega^4 \\ &- (\omega_M + \alpha^2\omega_M + 2\alpha\tau\omega_H\omega_M)\omega^2 \\ &+ \omega_H^2\omega_M\big],\end{aligned} \quad \text{(S9)}$$

$$\chi_a'' = \frac{2\alpha\omega_M\omega^2(\omega_H - \alpha\tau\omega^2)}{N}, \quad \text{(S10)}$$

$$\begin{aligned}N = &\ \alpha^4\tau^4\omega^8 \\ &+ 2\alpha^2\tau^2(-1+\alpha^2-2\alpha\tau\omega_H)\omega^6 \\ &+ (1+4\alpha\tau\omega_H+2\alpha^2+6\alpha^2\tau^2\omega_H^2 \\ &\quad -4\alpha^3\tau\omega_H+\alpha^4)\omega^4 \\ &+ 2\omega_H^2(-1+\alpha^2-2\alpha\tau\omega_H)\omega^2 \\ &+ \omega_H^4\end{aligned} \quad \text{(S11)}$$

The relation between $\chi_\pm$ and $\chi$, $\chi_a$ is elucidated in Fig. S1(a-d). Note that the ferromagnetic resonance (FMR) is observed for $\chi_+ = \chi + \chi_a$, whereas nutation resonance appears for $\chi_- = \chi - \chi_a$. Hence one can consider the sum and the difference of the $\chi$ and $\chi_a$ curves. The frequency dependence of $\chi_a$ in the terahertz range (Fig. S1(b)) is inverse to $\chi$ (Fig. S1(d)), and if one takes sum of the curves $\chi$ and $\chi_a$, then FMR peak increases in the microwave range, but nutation peak vanishes in terahertz range. One can perform a similar consideration for $\chi_- = \chi - \chi_a$.